\documentclass[reprint,aps,prb]{revtex4-1}

\usepackage[english]{babel}

\usepackage{graphicx}
\usepackage{subfig}
\usepackage{amsmath,amsfonts,amssymb,amsthm}
\usepackage{mathtools,braket}
\usepackage{tikz}
\usepackage{xifthen}
\usepackage{url}

\newcommand{\dd}{{\rm d}}
\renewcommand{\r}{{\bf r}}

\renewcommand{\AA}{{Angstr\"om}}

\newcommand{\eps}{\epsilon}

\newcommand{\ii}{\mathrm{i}}

\newcommand{\sint}{\mathrlap{\displaystyle\int}
\mathrlap{\textstyle\sum}
\phantom{\mathrlap{\displaystyle
\int}\textstyle\sum}}

\newcommand{\be}{\begin{equation}}
\newcommand{\ee}{\end{equation}}
\newcommand{\ba}{\begin{eqnarray}}
\newcommand{\ea}{\end{eqnarray}}
\newcommand{\baa}{\begin{align}}
\newcommand{\eaa}{\end{align}}
\newcommand{\nn}{\notag}
\newcommand{\qq}{\qquad}
\newcommand{\lb}{\label}
\newcommand{\mat}[1]{\begin{pmatrix} #1\end{pmatrix}}

\newcommand{\op}[1]{\hat {#1}}
\newcommand{\sop}[1]{\op{\op {#1}}}
\newcommand{\commutator}[2]{\left[ {#1} , {#2} \right]}
\newcommand{\trace}[1]{\mathrm{tr}\left(#1\right)}
\newcommand{\argument}[1]{\ifthenelse{\isempty{#1}{}}{}{(#1)}}

\newcommand{\optr}[1]{\check #1}

\newcommand{\brket}[2]{\langle  #1 | #2 \rangle} 

\newcommand{\ketbra}[2]{| #1 \rangle \langle #2 |}

\newcommand{\dmnot}{\op{\rho}_0}
\newcommand{\dm}{\op{\rho}}
\newcommand{\hnot}{\op{H}_0}

\newcommand{\excite}[2]{\op e_{{#1}{#2}}}
\newcommand{\decay}[2]{\op d_{{#1}{#2}}}
\newcommand{\Liouv}{\sop{\mathcal L}}
\newcommand{\Liouvnot}{\sop{\mathcal L_0}}
\newcommand{\coupl}{\sop{\mathcal K}}

\newcommand{\identity}{\op{\mathbb I}}
\newcommand{\rmat}[1]{\optr R}

\newcommand{\GH}{\op G_{\op T}}

\begin{document}


\title{Locality and Computational Reliability of Linear Response Calculations for Molecular Systems}
\author{Marco D'Alessandro}
\affiliation{Istituto di Struttura della Materia-CNR (ISM-CNR), Via del Fosso del Cavaliere 100, 00133 Roma, Italia}
\author{Luigi Genovese}
\affiliation{Laboratoire de Simulation Atomistique (LSim), Univ. Grenoble Alpes, CEA, INAC-MEM, 38000 Grenoble, France}
\date{\today}

\begin{abstract}
By performing a critical analysis of the fundamental equations of linear-response (LR) formalism in molecules,
we explore the interplay between locality of the response density operator and  numerical convergence of LR-related quantities.
We show that for frequencies below the first ionization potential (IP) of the system, it is possible to express the response density by employing localized states only.
Above this threshold energy, such a locality property cannot be achieved.
Such considerations may be transposed in terms of the molecule's excited states. We show that not all the system's excitations can be considered on equal footing.
There is a discrete sector of excitations -- which may also extend above IP -- that can be parametrized by observable, localized states, which can be computationally expressed with high precision, provided an adequate level of completeness.
We present indicators that can help to quantify such potential observable properties of an excitation, that can be evaluated
in any discretization scheme.
The remaining excitation modes belong to a continuum spectrum that, on the contrary, is not directly associated to observable properties and can only be effectively represented in a given computational setup.
Such considerations are important not only for reproducibility of the results among different computer codes
employing diverse formalisms, but also in view of providing a deeper understanding on the impact of models'
approximations on the scientific outcomes of the simulation.
\end{abstract}

\maketitle

\section{Introduction}

For a theoretical approach that is based on numerical calculations, for which no analytic reference solution exists,
reducing the computational uncertainty is the only possible way to shed light on the predictive power of the
model. In other terms, the ``accuracy'' of a result with respect to experimental data may be reliably quantified
only when the computational uncertainty is guaranteed to be significantly lower than the observed discrepancy.
For communities employing density functional theory (DFT), the need of such a ``calibration'' of computer codes has led to the DELTACODES project~\cite{deltaTest2016}.
This project represents a remarkable example of precision-driven initiative where a grassroots community of code-owners struggled to extract
well-defined computational results in the context of DFT calculations. Ground-state quantum mechanical quantities
as lattice constants and bulk modulus of elementary crystals extracted from different codes were compared with each other, with the objective of reducing the uncertainties on such data, and in some sense, to define a \emph{calibration} procedure for a solid-state Physics DFT code.
Any other code able to extract the same quantities with the same approximations may be compared with the
results of the community in order to assess its computational reliability.

The problem is even more stringent for calculations that refer to quantities \emph{beyond} ground state, like
for instance the time-dependent density functional theory (TDDFT) \cite{casida1995,runge1984,onida2002},
from which quantities like optical excitations or polarizability tensors
of a given system may be studied and put in relation with experimental data.
Here also, a plethora of computational approaches exist, that employ different numerical formalism and
basis sets, and it will be of great importance to identify quantities that can be ``measured'' in a given model, to assess
and quantify their computational uncertainty.

In this paper, we focus our attention on linear-response (LR) treatments of \emph{molecular} systems,
in particular referring to LR-TDDFT calculation. We will investigate the impact that \emph{open} (i.e., isolated) boundary conditions (BC), which are imposed to the system, will have on the computational evaluation of LR quantities.
We perform analytic analysis on the equations beyond LR treatment and discuss the interplay between reproducible LR quantities and the \emph{locality} of the related objects. Such considerations will enable us to distinguish quantities which \textit{a priori} can be compared among different treatments - and therefore tested for reproducibility - from others which would \emph{explicitly} depend of the employed computational treatment.

It is indeed relatively easy to have converged indicators for numerical ground-state (GS) quantities: on one side, the variational theorem
guarantees that the lower the GS energy, the more precise the numerical result; on the other side, GS quantities of molecular system  can be expressed only in terms of localized (i.e. bound) states. These considerations make it relatively easy to increase the basis set completeness in view of the extraction of reference results.

We will see in what follows that the situation is not as simple in the context of LR. Indeed the response density -- even for molecular systems -- loses its localized behavior for high frequencies, independently on the nature of the specific system and perturbation. Furthermore, we will demonstrate that to increase the basis set completeness, delocalized oscillatory degrees of freedom cannot be excluded \textit{a priori}, even in the localized regime.

Lastly, we will show that the reliability of the convergence criteria that can be established for LR, where no variational theorem exists -- depends explicitly on the target frequency range and on the features of the numerical basis set.

The paper is organized as follows.
Initially, we will inspect how the question of reproducibility may be expressed
for ground state calculations of molecules.
Following these guidelines, we will then move to LR equations,
by first considering the equations of motion of the response density operator, given
the application of a perturbing field.
Then, we will consider the behavior of the ``free oscillations'', i.e. the excited states
of the molecule.
Our considerations will prove to be useful for the understanding of the \emph{analytic structure} of LR quantities like
the linear susceptibility of the system (i.e. the reducible polarizability).
We support our considerations with numerical results on simple LR-TDDFT quantities.
For readability purposes, the details of each calculation are explained in Appendix~\ref{compdetails}.

Although we refer, as anticipated, to TDDFT models, such considerations might also be transposed to many-body perturbation theory
approaches, in the context of quasiparticle equations.

\subsection{Long range behavior of the eigenvalue solution of the Schr\"odinger equation}
\label{SEopenSystem}

Let us start our discussion by reviewing some well-known concepts of quantum mechanics under a perspective that will turn out to be very useful
in the forthcoming discussions about linear-response quantities. In particular we revisit the impact that the isolated BC have on the solutions
of the Schr\"odinger equation.

Consider a (one-body) wave function $\ket{\psi}$ solution of the equation
$\op H\ket{\psi}=\eps\ket{\psi}$, where we split the Hamiltonian operator into the sum of
a kinetic $\op T$ and a potential $\op V \equiv \op H - \op T$ term.
In view of discussing the impact of the BC, it is instructive to transform
the Schr\"odinger equation in a scattering problem, by writing
\be\lb{SEHelmholtz1}
\ket{\psi} = \GH(\eps)\ket{\op V\psi}\;,
\ee
where we have introduced the Green function of the Helmholtz operator $\GH(\eps) = (\eps-\op T)^{-1}$,
resolvent of the free-particle Hamiltonian $H_s \equiv \op T$.

When describing the ground state of a molecule, we suppose that the potential operator, even if non-local,
can be effectively restricted to wave functions that can be considered as nonzero only on a
bounded domain. In other terms, a molecule can be associated to a ``scatterer'' that has a \emph{finite}
spatial extension; if a state $\ket{\psi}$ is localized, it will be the same for the ket $\ket{\op V \psi}$, though
perhaps in a larger region.
For a computational analysis of the problem, this property is fundamental to the choice of the numerical treatment as we know that
computational basis sets
which are tailored to express asymptotically vanishing wave functions are better suited for the solution of the problem.

This fact is made apparent by the expression of
the kernel of $\GH(\eps)$ in the coordinate representation.
When dealing with Isolated BC, it is easy to see that the solution of Eq.~\eqref{SEHelmholtz1}
can be expressed by:
\be\lb{HelmholtzKernelDef1}
\bra{\r} \GH(\eps) \ket{\r'} = \frac{1}{4\pi} \begin{cases}
\frac{e^{-\alpha|\r - \r' |}}{|\r- \r' |} \;,\; & {\textrm for}\; \eps  < 0\; \\
\frac{e^{\ii \alpha |\r-\r' |}}{|\r-\r' |} \;, & {\textrm for} \; \eps \geq 0
\end{cases} \;,
\ee
where $\alpha = \sqrt{2|\eps|}$.

From the above expressions we can see that, for molecular systems,the value $\epsilon=0$
represents a \emph{threshold}~\footnote{The threshold value in general should correspond to asymptotic value of the potential at infinity.}
energy that separates two different classes of solutions.
Negative-energy solutions are \emph{localized}, i.e. they exhibit bound-state behavior far away from the scatterer.
These are bound states and, by means of their normalizability, they are associated to \emph{discrete}, well-defined energies.
The bound states can be of course labeled by quantum numbers which are defined by the properties of the scatterer, namely the
symmetries of the potential $\op V$.
In a computational discretization of the Schr\"odinger equation, such states can in principle be expressed with arbitrary precision,
provided that the numerical treatment offers an adequate level of completeness.
Being discrete and well-identified, the bound-state energies are a property of the molecule and can be associated to \emph{observable}
quantities. Their energies can be compared among different treatments and it is in principle possible to
provide reference values.

On the other hand, positive energy solutions behave differently. Even though the potential is localized the associated Helmholtz kernel is a spherical wave and its convolution with the source term provides a delocalized wavefunction $\psi(\r)$ in the whole space.
As they cannot be normalized they belong to the essential spectrum of the Hamiltonian operator and form a \emph{continuum} of states.
Outside of the support of the $\ket{\op V \psi}$ the solutions of Eq.~\eqref{SEHelmholtz1} can be put in bijection with the eigenstates of the
free-particle Hamiltonian $\op H_s$. For positive energies, the quantum numbers of the eigenstates are determined by the Laplacian operator,
regardless of the particular features of the potential.

It is easy to see from these concepts that a computational treatment that is tailored for the discretization of bound states
will not be as effective for continuum states, due to the different long-range asymptotic behavior - which in turn is a clear consequence of the
BC of the problem. In a numerical treatment, where we discretize the problem with a \emph{finite}
number of degrees of freedom, we will only have access to a \emph{pseudo}-continuum of states,
whose energy values and density of states will depend on the representation of the kinetic operator
in the basis set. As a consequence, the set of eigenstates with $\epsilon > 0$ expressed in a given computational setup depends on the features of the employed basis set.

Direct numerical evidence of this problem has been discussed in \cite{boffi2016}, where the authors analyze the effects of a finite basis set on the
occupied and virtual orbitals of a molecular system and establish a clear correspondence between the orbital localization character,
and the independence of its energy with respect to the basis set.
Furthermore, it is clearly shown that the energies of the unbound orbitals are not stable with respect to parameters like the size of the computational domain.

\section{Fluctuation states. A localization argument for the response density}
\label{FluctuationState}
It is evident that the arguments presented above have a limited interest for the study
of ground-state quantities of molecules. By definition, all these quantities can be expressed as a
functional which \emph{entirely} depends on bound states. As an illustration it is sufficient to recall
the Hohenberg-Kohn theorem that states that the ground-state (GS) energy is a functional of the GS
charge density $\dmnot$. For molecular systems which are electronically stable (i.e. no anionic or metastable configurations),
such a quantity is localized, in the sense defined in the above section. Stated otherwise,
\emph{there is no need to express efficiently any portion of the continuum spectrum} to have a reliable GS treatment of molecules.

Let us now inspect the case of LR calculations.
In this framework, we consider the unperturbed Hamiltonian $\hnot$ and assumes that
the GS density $\dmnot$ is accessible and expressed in a given computational treatment.
The linear response formalism allows us to evaluate the modification of the expectation value of a generic observable induced by a
time- (or frequency-) dependent perturbing field $\op\Phi(\omega)$ acting on the system.
This quantity, written in the frequency domain, can be expressed through the evaluation of the \emph{linear response functional}
\be\lb{LinearResponseFunctDef1}
\braket{\delta\op O}_\Phi(\omega) = \trace{\dm'_\Phi(\omega)\op O} \;,
\ee
where $\op O$ represents the observable\footnote{For a sake of concreteness, we are limiting our considerations to observables that do not explicitly depend on time} under inspection
and we have introduced the \emph{response density operator} $\dm'_\Phi(\omega)$ that codifies the modification of $\dmnot$ induced by the perturbation $\op\Phi(\omega)$.

The response density operator is expressed as the first-order variation of $\dmnot=\sum_{\{p\}} \ket{\psi_p} \bra{\psi_p}$, where the set $\{\ket{\psi_p}\}$ denote the
occupied states, \emph{i.e} the subset of the bound eigenstates\footnote{In what follows we will assume that the set of states $\{\ket{\psi_p}\}$ are described by real
functions, when projected in the $\r$-representation.} of $\hnot$ with (negative) energy $\eps_p$ lower than the Fermi level. Using this notation we denote the energy of the HOMO level
as $\eps_h$ and identify the value of the first ionization potential with IP~$\equiv|\eps_h|$.

The response density satisfies an equation of motion written in the form of a quantum Liouville operator (for example see \cite{baroni2008})
\be\lb{LiouvillianRhopomegaDef1}
\left(\omega - \Liouv\right) \dm'_\Phi(\omega) =  \commutator{\op\Phi(\omega)}{\dmnot} \;.
\ee
The Liouvillian superoperator $\Liouv$ for self-consistent systems is expressed as the sum of the unperturbed part $\Liouvnot$ plus a coupling term $ \coupl$. The action of these
terms reads, respectively
\be\lb{LiouZeroDef1}
\Liouv \op O \equiv \left(\Liouvnot + \coupl \right) \op O = \commutator{\hnot}{\op O} +\commutator{\op V'[\op O]}{\dmnot}  \;,
\ee
where $\op V'[\op O] \equiv \int \dd \r \dd \r'\frac{\delta \op V[\dmnot]}{\delta \rho(\r,\r')} O(\r,\r')$ encodes the response of the $\dm$-dependent potential to a modification of the density operator.
The inspection of the linear order time evolution of the eigenstates of $\hnot$ shows that the response density acts as a transition
operator, linking the occupied and empty subspaces of $\hnot$.
An operator $\op O_\perp$ with this feature satisfies the \emph{transverse} condition
\be\lb{RhopTransverseDef1}
\op O_\perp \equiv
\dmnot\op O \op Q_0 + \op Q_0 \op O\dmnot \;,
\ee
where $\op Q_0=\identity-\dmnot$ is the projector to the empty subspace of $\hnot$
(for more details see Appendix~\ref{LiouvillianAction}).
In other terms, the condition $\dm'_\Phi = \dm'_{\Phi\,\perp}$ holds.

We can therefore introduce an explicit representation of the
response density parametrized as
\be\lb{rhoPrimeFluctuationStateDef1}
\dm'_\Phi(\omega) = \sum_p\left(\ketbra{\psi_p}{f_p^\Phi(-\omega)} + \ketbra{f_p^\Phi(\omega)}{\psi_p}\right) \;.
\ee
Here we have introduced the set of $\omega$-dependent \emph{fluctuation states} (FS)
$\ket{f_p^\Phi(\omega)} = \dm'_\Phi(\omega) \ket{\psi_p}$, that have values in the unoccupied subspace of $\hnot$.
However, non-Hermiticity of $\dm'_\Phi(\omega)$ implies that the
fluctuation states are complex quantities.

An analysis of the Liouville equation \eqref{LiouvillianRhopomegaDef1} for the response density written in this fashion evidences that the equations of motion of fluctuation states  are
written as a modified Sternheimer equation~\cite{mahan1980,giustino2012,giustino2014}
\be\lb{fluctuationStateEqMotion1}
\left[\omega - (\hnot-\eps_p)\right]\ket{f_p^\Phi(\omega)} = \op Q_0(\op\Phi(\omega)+\op V'[\dm'_\Phi](\omega))\ket{\psi_p} \;,
\ee
and we observe that the first-order Hamiltonian contains,
apart from the perturbing field, a further term $\op V'[\dm'_\Phi]$ due to the density-dependence of $\hnot$.
It is easy to see that the solution of Eq. \eqref{fluctuationStateEqMotion1} can be recast in a scattering problem:
\be\lb{rhoPrimeFluctuationStateDef2}
\ket{f_p^\Phi(\omega)} = \GH(\omega+\epsilon_p)\left(\op V\ket{f_p^\Phi(\omega)} + \ket{s_p^\Phi(\omega)} \right)\;,
\ee
with the state
$
\ket{s_p^\Phi(\omega)} = \op Q_0(\op \Phi(\omega) +\op V'[\dm'_\Phi](\omega) )\ket{\psi_p}
$ as a source term.

From equation \eqref{rhoPrimeFluctuationStateDef2} we see that for each choice of $p$, the value $\omega = |\eps_p|$ represents the threshold level that governs the asymptotic behavior of the associated Helmholtz kernel.
As a ulterior \emph{stringent} hypothesis, let us now assume that the source term $\ket{s_p^\Phi(\omega)}$ is spatially localized, \textit{i.e.} its evaluation can be restricted to a bounded real-space domain.
Even when such a locality argument holds true, for $\omega > |\eps_p|$,
the Helmholtz kernel behaves as a spherical wave and  gives rise to a $\ket{f^\Phi_p(\omega)}$ which is \emph{delocalized} over the entire real-space domain.
Such delocalization is an intrinsic consequence of the value of $\omega$, and does not depend on the particular choice of
$\op \Phi$.
Only for $\omega < |\eps_p|$ the kernel has an exponential damping factor and $\ket{f^\Phi_p(\omega)}$ exhibits a bound-state behavior in the long range,
once again provided that the above locality arguments on $\ket{s_p^\Phi(\omega)}$ -- hence on $\op V'$ and $\Phi$ -- are valid.

These arguments demonstrate that for frequencies above the IP threshold energy, the response density operator cannot be expressed in terms of localized states. Locality of FS (and therefore of $\dm'_\Phi(\omega)$) can be only recovered for frequencies below such threshold.

\subsection{Localized behavior for static perturbations}
The locality property of the fluctuation state can be directly confirmed for static perturbations at $\omega=0$.
Let us consider a static perturbing operator $\op\Phi_s$ that can be restricted to a finite domain.
Any local operator, \emph{e.g.} a static electric field $\mathbf F$, with associated scalar potential $\Phi_s[\mathbf F](\r) = -\mathbf F \cdot \r$, would satisfy such a property.
We consider the perturbed Hamiltonian $\hnot[\dm_{\Phi_s}] + \op\Phi_s[\mathbf F]$, where $\dm_{\Phi_s}$ represents the GS density of the perturbed system, and assume that the
perturbation strength ($|\mathbf F|$ in the above example) is sufficiently small to be consistent with a LR treatment of the problem.
In this case, the occupied eigenfunction $\ket{\psi_p^{\Phi_s}}$ of the perturbed Hamiltonian can be extracted with traditional GS techniques,
since from the discussion around Eq.~\eqref{SEHelmholtz1}, they exhibit bound-state behavior.
\begin{figure}[t]
\includegraphics[scale=0.68]{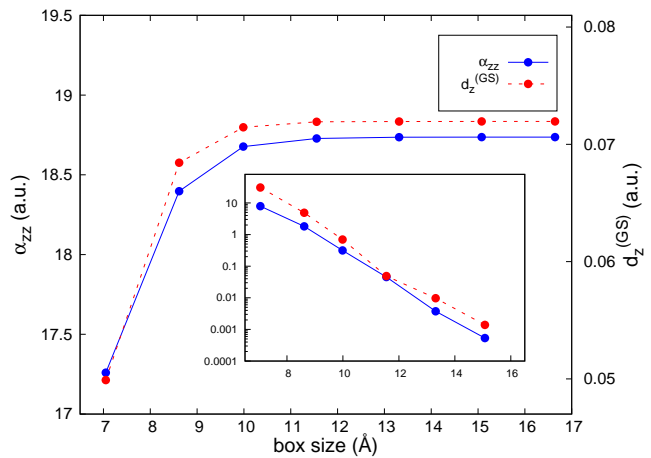}
\caption{\label{co_alphaStatic}(Color online) Convergence of the static polarizability $\alpha_{zz}$ (blue continuous line) and of the ground state dipole $d_z$ (red dotted line)
of a $CO$ molecule as a function of the size of the computational domain. The inset evidences the relative convergence (in percent) of the two quantities with respect to the corresponding value associated with the largest
box.}
\end{figure}

From the linearized Schr\"odinger equation of the perturbed problem, it is easy to see that the FS at $\omega=0$ can be expressed
in terms of the corresponding bound state of the perturbed Hamiltonian, i.e.
$\ket{f_p^{\Phi_s}(0)} \simeq \op Q_0 \ket{\psi_p^{\Phi_s}}$.
For any value of $p$, the locality of the $\ket{f_p^{\Phi_s}(0)}$ therefore directly stems from the
bound-state behavior of $\ket{\psi_p^{\Phi_s}}$. As $\omega=0$ is by hypothesis
lower than any of the $|\eps_p|$, this is consistent with the above considerations.

Thanks to such observations one can express quantities like the static polarizability tensor $\alpha_{ij}$ (see e.g.\cite{DebElecField}) from GS calculations of the perturbed Hamiltonian.
By expressing the induced dipole $\braket{\delta \op{\mathbf r}}_{\Phi_s}$ in the form Eq.~\eqref{LinearResponseFunctDef1}, and identifying the static response density as $\dm'_{\Phi_s}(0) = \dm_{\Phi_s} -\dmnot $, we have
\be \label{staticalpha}
\alpha_{ij} =
-\frac{1}{F_j} \left(\trace{\op r_i \dm_{\Phi_s[F_j]}} - d^{\text{(GS)}}_i \right)\;,
\ee
where $\mathbf d^\text{(GS)}=\braket{\mathbf r}$ is the ground state dipole.
Figure~\ref{co_alphaStatic} illustrates this procedure for a $CO$ molecule
perturbed by a static electric field.
Care has been taken in keeping the field strengths $F_j$ small enough to
preserve the validity of the linear response regime.
The local character of the fluctuation states at $\omega=0$ is
probed by verifying the convergence of the statical polarizability versus
the size of the computational domain used to express the fluctuation states. The convergence rate is compared with the
one of the static dipole of the molecule computed at zero external field.
The results show that the typical sizes of $\ket{f^\Phi_p(0)}$ are analogous to those of the unperturbed KS orbitals, which is a direct proof of the bound-state behavior of the FS.

\subsection{Localization properties at finite $\omega$. Emergence of a strong and a weak regime}

The analysis described above has shown that the building blocks of the response density $\{\ket{f_p(\omega)}\}$
exhibit bound-state long-range behavior \emph{only} when $\omega < |\eps_p|$,
while they behave as unbound states for values of $\omega$ above these threshold levels.
This fact has interesting implications in the evaluation of LR quantities. Indeed, writing the response
density in terms of FS implies that
\be\lb{LinearResponseFunctDef2}
\braket{\delta\op O}_\Phi(\omega) = 2 \sum_p\bra{\psi_p}\op O\ket{f_p^\Phi(\omega)}  \;,
\ee
where we have assumed that $\op O$ is a real symmetric operator and chosen a real set of occupied molecular states.
Let us suppose that the observable of interest $\op O$ also can be restricted to localized domains in the same way as the potentials $ \op V$ and $\op \Phi$.
Once again, it would be enough to consider a local operator for this condition to be met.

Two distinct regimes can be identified, on the basis of the value of $\omega$.
The first one, which we will refer to as the \emph{below threshold} regime, is realized when
$\omega$ is lower than the ionization potential $|\eps_h|$.
In this case all the FS are below their threshold level and Eq. \eqref{LinearResponseFunctDef2}
is expressed in terms of genuinely localized
quantities. At these frequencies, the linear response functional $\braket{\delta\op O}_\Phi(\omega)$
may thus be evaluated with computational setups that
are similar to those usually employed for GS calculations.
The convergence in this regime is ``strong'', in the sense that the basis functions employed enable us to express
with arbitrary precision \emph{both} the states $\ket{\op O \psi_p}$ and $\ket{f_p^\Phi(\omega)}$.
This is, of course, true even when the above states are only implicitly expressed by the employed LR treatment, as e.g. in the case of Krylov spaces generated from the states $\ket{\op\Phi(\omega) \psi_p}$ \cite{baroni2006,baroni2008,linlinKPM}.
In this regime, a computational treatment based on localized basis set is susceptible to provide a precise answer, assuming a reasonable level of completeness.

On the contrary, the computation of the linear response functional can be much more demanding in the \emph{above threshold regime},
realized for $\omega>|\eps_h|$, in which fluctuation states start behaving as unbound wave functions.
As by hypothesis $\ket{\op O \psi_p}$ is bound-state like, \emph{only} the scalar products of Eq.~\eqref{LinearResponseFunctDef2}
can be evaluated in a localized domain. In this case a computational description of fluctuation states is only meaningful in the ``weak''
sense, where the bound state character of $\bra{\psi_p\op O}$ behaves as a regulator. This is naturally achieved in the above mentioned
Krylov space treatments, which explicitly deal with the expression of the scalar products. However, in this regime, imposing \emph{by design}
a localized behavior of the FS might reveal very dangerous in view of convergence of the results since the computational basis set has
to be able to express a delocalized FS in the support domain of $\bra{\psi_p\op O}$ .
This fact implies that the degrees of completeness of the computational basis should be increased by adding
the oscillatory degrees of freedom needed to describe the FS in the relevant domain.
Stated otherwise, to achieve precise results in this regime the
underlying basis set has to be able to express delocalized states.

These considerations prove why, when a given computer code employs established GS numerical techniques,
\emph{it is much easier to converge LR quantities for $\omega$ below IP}.
Above the IP threshold, a computational setup
provides a reliable assessment of \eqref{LinearResponseFunctDef2} only if it is able (even implicitly) to express an
unbiased description of the fluctuation states in the support of $\{\bra{\psi_p}\op O\}$.
The non-vanishing spherical-wave behavior in the long range of $\ket{f_p^\Phi(\omega)}$ is therefore responsible
for these convergence difficulties, especially for high frequencies.
The IP threshold has an evident physical interpretation: for energies in a range higher than the ionization potential,
localized and delocalized states might be coupled by the perturbation. Obtaining convergence of the observable quantities at these energies is therefore much more challenging if one employs
numerical techniques which were originally tailored for GS calculations.

\begin{figure}
\centering
\subfloat
{\includegraphics[scale=0.56]{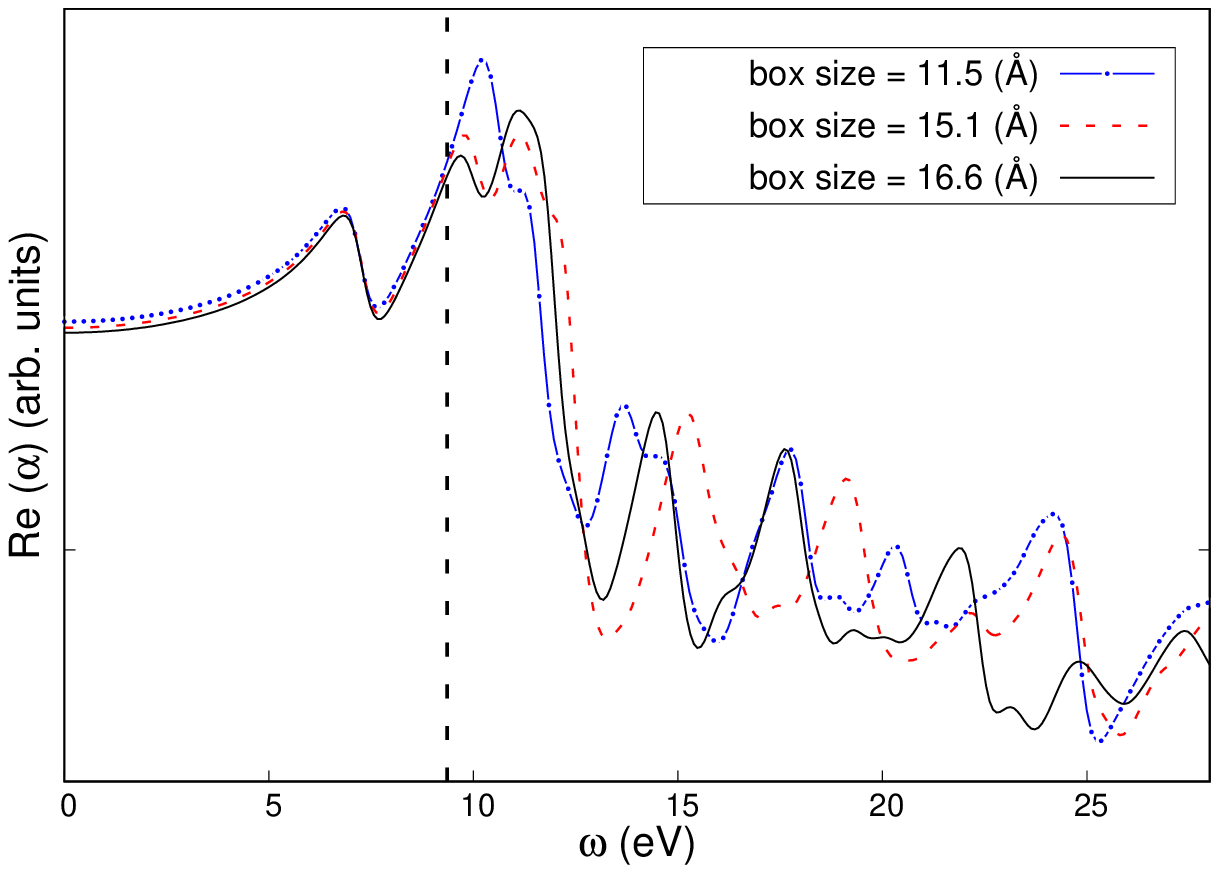}} \\
\centering
\subfloat
{\includegraphics[scale=0.58]{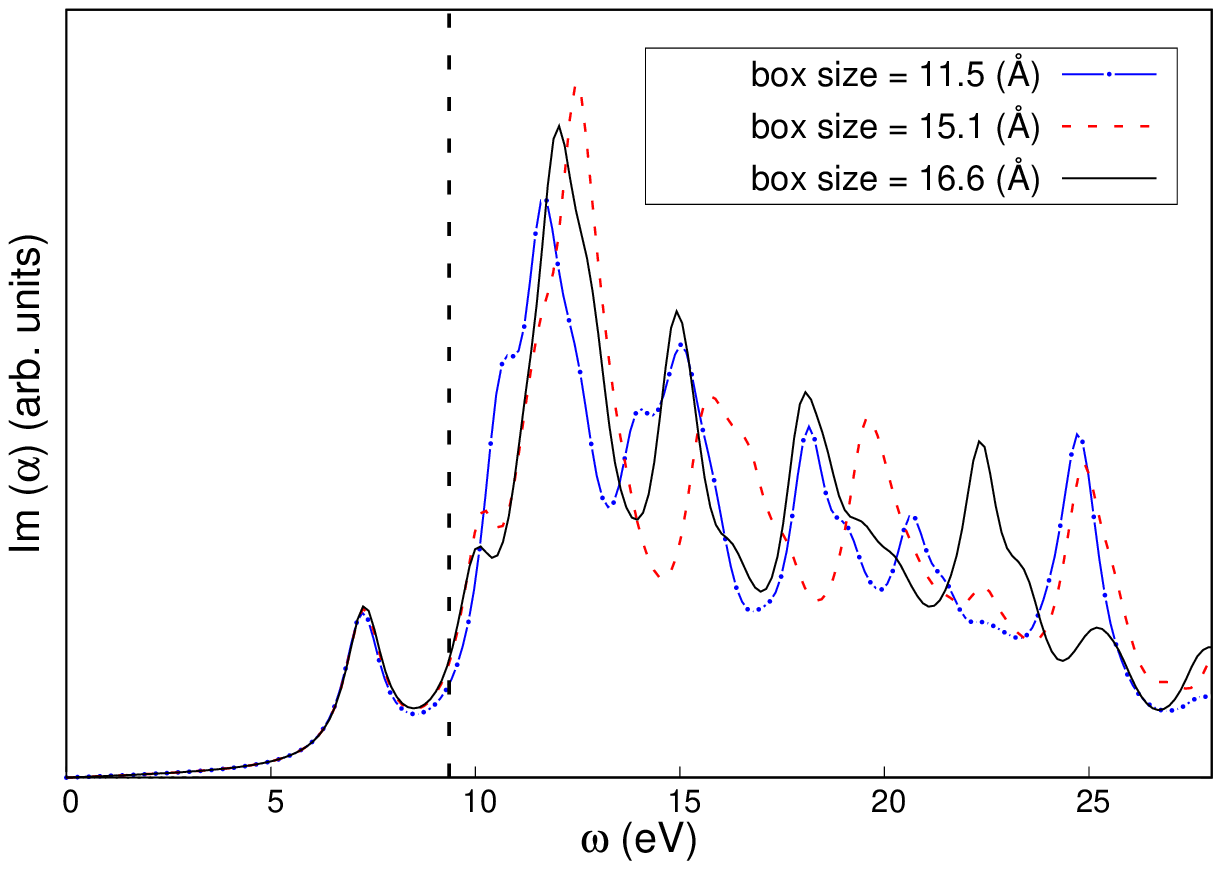}}
\caption{\label{co_spectrum}(Color online) LR dynamical polarizability of the $CO$ molecule in computational setups with different
sizes of the simulation domain. The top panel displays the real part and the bottom one the imaginary part. The value of the IP threshold
is indicated by a vertical dashed line.}
\end{figure}

As an example in support of these arguments we compare the {dynamical polarizability $\alpha(\omega)$} of a $CO$ molecule computed in three distinct real-space computational setups, which differ in the dimension of the computational domain.
Results reported in the two panels of Fig. \ref{co_spectrum} show the real and imaginary parts of $\alpha(\omega)$, respectively. In all the cases, a typical behavior emerges: the curves are almost coincident in the below threshold regime,
whereas the effects of the computational setup are clearly visible for values of  $\omega$ which exceed the threshold value.
For these high-frequency regions, convergence might be reached for domains of much larger sizes than those presented here~\cite{baroni2008},
which are, as already noticed, more than sufficient below threshold.

\section{Excitation operators. The analytic structure of the susceptibility functional}

It is possible to reformulate this type of analysis by directly looking at LR quantities that are defined independently from the perturbing operators and target observable. This can be achieved by studying the \emph{susceptibility functional}, defined implicitly through the operator equation
\be\lb{SusceptibilityDef1}
\dm'(\omega) = \left(\omega -\mathcal L \right)^{-1} \commutator{\op \Phi(\omega)}{\dmnot} = \sop{{\cal \chi}}(\omega) \op\Phi(\omega) \;.
\ee
The susceptibility (super)operator $\cal \chi(\omega)$ is therefore closely related to the resolvent of the Liouvillian of Eq.~\eqref{LiouvillianRhopomegaDef1}, which is completely defined
from the molecule Hamiltonian $\hnot[\dm]$.
The ``excitation modes''  of the molecule are defined through the \emph{excitation operators}, satisfying
\be\lb{ExcitationOperatorsDef1}
\Liouv \op E_a = \Omega_a \op E_a \qq \op{\tilde E}_a \Liouv = \Omega_a \op{\tilde E}_a \;,
\ee
together with the operator orthonormalization condition 
\be\lb{orthoExcitatioOpDef1}
\trace{\op{\tilde E}_a\op E_b} = \delta_{ab} \;.
\ee
Excitations defined in this way may be considered as a basis of the operators that satisfy the transverse condition as in
Eq.~\eqref{RhopTransverseDef1}, and enable us to express the susceptibility as a spectral decomposition
\be\lb{RhopExcitationDef1}
\sop{{\cal \chi}}(\omega) = \sum_{\{a\}} \,
\frac{\sop B_a}{\omega-\Omega_a} \;,
\ee
where the action of the spectral (super)operator $\sop B_a$ reads
\be\lb{spectralop}
  \sop B_a \cdot = \op E_a \trace{\commutator{\dmnot}{\op{\tilde E}_a} \cdot} \;. \nn
\ee
These formulas allow us to present an excitation-based description of Eq.~\eqref{LinearResponseFunctDef2} as follows
\be\lb{deltaoexc}
  \braket{\delta \op O} =  \sum_{\{a\}} \frac{\trace{\op O \sop B_a \op \Phi(\omega)}}{\omega -\Omega_a}\;.
\ee

\subsection{Localization features of the excitation operators. The analytic structure of linear susceptibility}

The properties of the Liouvillian superoperator $\Liouv$ (sketched in Appendix \ref{LiouvillianAction}) imply that both the left and right operator-valued
eigenstates \eqref{ExcitationOperatorsDef1} satisfy the transverse condition \eqref{RhopTransverseDef1} and can be parametrized as
\begin{align}\lb{ExcitationOperatorsDef2}
\op E_a &= \sum_p \left( \ket{\phi^a_p}\bra{\psi_p} + \ket{\psi_p} \bra{\chi^a_p}\right)\;, \nn \\
\op{\tilde E}_a = \commutator{\dmnot}{\op E^t_a} &= \sum_p \left( \ket{\psi_p}\bra{\phi^a_p} - \ket{\chi^a_p} \bra{\psi_p}\right)\;.
\end{align}
Each excitation ``mode'' of the system, with associated energy $\Omega_a$, is thus described by a set of states $\{\phi^a_p,\chi^a_p\}_{p \in a}$, defined in the unoccupied subspace. These objects represent, respectively, the state in which $\ket{\psi_p}$ is excited -- or from whom it decays -- when the system is subject to the \emph{monochromatic} perturbation $\op \Phi_a \equiv \commutator{\op E_a}{\dmnot}$. We denote with the shorthand $p \in a$ the fact that the sum of Eq.~\eqref{ExcitationOperatorsDef2} is in general restricted only to a $a$-dependent subset of the occupied orbitals, depending on the particular symmetry
of the excitation. Excited states  satisfy the normalization condition
\be\lb{ExcitedStateOrthNormDef1}
\sum_p \left(\brket{\phi_p^a}{\phi_p^b} - \brket{\chi_p^b}{\chi_p^a}\right) = \delta_{ab} \;.
\ee
By applying \eqref{ExcitationOperatorsDef1} to the above parametrization
of the excitations, we obtain the following Sternheimer-type equations for the excited states:
\begin{align}\lb{ExcitationOperatorsDef3}
&\left[\Omega_a-(\hnot - \eps_p)\right] \ket{\phi_p^a} = \op Q_0 \op V'[\op E_a] \ket{\psi_p} \;, \nn\\
&\bra{\chi_p^a}\left[-\Omega_a-(\hnot - \eps_p)\right] = \bra{\psi_p} \op V'[\op E_a] \op Q_0  \;.
\end{align}
By writing these equations in a basis set of virtual states we obtain the well-known Casida's eigenvalue equation\cite{casida1995}
associated to the excitation energy $\Omega_a$ (see Appendix \ref{casida}).
The spectrum is symmetric with respect to the inversion of the eigenvalues $\Omega_a \rightarrow -\Omega_a$ and, given a specific excitation $\{\phi^a_p,\chi^a_p\}$,
the associated solution with opposite energy is described by the transposed pair $\{\chi^a_p,\phi^a_p\}$.
We concentrate our analysis on the sector of positive energies, i.e. $\Omega_a > 0$.

In the same spirit as the previous sections, we present a formal solution of \eqref{ExcitationOperatorsDef3} as follows
\begin{align}
\ket{\phi^a_p} &= \GH(\Omega_a+\eps_p)\left(\op V \ket{\phi^a_p} + \ket{s_p[\op E_a]} \right) \;, \nn \\
\bra {\chi^a_p} &=\left(\bra{\chi^a_p} \op V  + \bra{s_p[\op E_a]}\right)  \GH(-\Omega_a+\eps_p)\;,
\end{align}
where the source terms are here defined as:
\be
 \ket{s_p[\op O]} =  \op Q_0 \op V'[\op O] \ket{\psi_p}\,, \quad
 \bra{s_p[\op O]} =   \bra{\psi_p} \op V'[\op O] \op Q_0 \nn \;.
\ee
First, we point out that the Helmholtz kernel associated with the $\ket{\chi^a_p}$ state
contains an exponential damping factor for \emph{all} the positive values of $\Omega_a$.
The asymptotic long-range behavior of $\ket{\chi^a_p}$ thus depends only on the locality properties of
$\bra{s_p[\op E_a]}$, which in turn are related to the behavior of the operator $\op V'[\op E_a]$.

On the other hand, the long-range behavior of $\GH(\Omega_a+\eps_p)$ depends on
$\Omega_a$, with a threshold of value $\Omega_a=|\eps_p|$.
For energies above this value, the behavior of the state $\ket{\phi^a_p}$ is
delocalized regardless of the particular nature of the operators $\op V$ and $\op V'$.
This means that, if an excitation operator $\op E_a$ has an energy such that
$\Omega_a>|\eps_p|$  for at least one of the $p \in a$ defining the excitation,
such an operator \emph{cannot} be parametrized by employing localized states only.

As a consequence, we deduce that a molecular excitation \emph{may} be expressed in terms of localized states
\emph{only} when its energy is lower than \emph{all} the ionization potentials of the occupied states participating to the excitation,
namely, if $\Omega_a<|\eps_p|$ $\forall p \in a$. This will only be true if the operator
$\op V'[\op E_a]$ can be restricted to localized kets and bras.
If any of these hypotheses does not hold, the excitations of the systems are defined via genuinely delocalized wave functions.
Furthermore, excited states undergo the orthonormalization condition
\eqref{ExcitedStateOrthNormDef1} that acts as an ulterior constraint and has the effect of determining the \emph{quantization}
of the energy levels $\Omega_a$ associated to the bound-state-like objects.
For delocalized states, as in the case of the continuum states of $\hnot$,
such a condition has to be interpreted in a distributional sense and does not lead to quantization of $\Omega_a$.
Thus we can conclude that all the states $\ket{\chi^a_p}$ and the $\ket{\phi^a_p}$ with $\Omega_a$ below the threshold value constitute
-- if localized -- a genuine \emph{discrete} set, while those in the unbound energy range behave as generalized eigenvectors associated to a
continuum spectrum.

\begin{figure}[!t]
\includegraphics[scale=0.12]{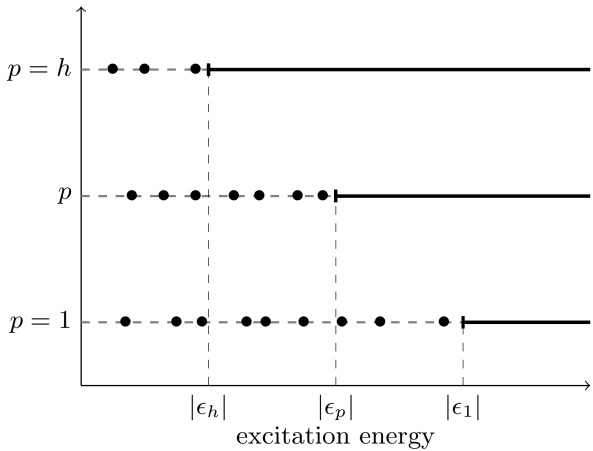}
\caption{\label{ExcitationLandscape} Excitation landscape. Excitations are split in horizontal lines according to the threshold
levels $p$. Filled bullets refer to localized excitations whereas the thick black lines describe delocalized ones belonging a
continuum spectrum.}
\end{figure}

The susceptibility functional has therefore a peculiar analytic structure made of multiple threshold energies, one for each of the ionization potentials associated with the occupied states. If an excitation $\op E_a$ has an energy $\Omega_a < |\eps_a|$ with $\eps_a = \mathrm{max}\left(\eps_p\right)_{p\in a}$, it may belong to a discrete spectrum; otherwise it \emph{has} to belong to a continuum of states.
The typical structure of the excitation landscape is depicted in Fig.~\ref{ExcitationLandscape}, once again assuming that the perturbed potential $\op V'[\op E]$ gives rise to a bound state when applied to $\ket{\psi_p}$. Below the first ionization energy $|\eps_h|$ only localized excitations are present. Conversely, in the intermediate energy interval that extends from $|\eps_h|$ to the absolute value of the deepest occupied orbital, other discrete excitations may exist, and are embedded in a continuum of delocalized excitations. For energies higher than $|\eps_1|$ \emph{no localized} excitations are anymore possible,
regardless of the particular behavior of $\op V'$: the excitation landscape in this region contains a continuum set of delocalized states.

\section{Physical relevance of the excitation spectrum. A comparison among localized and continuum sectors}

\subsection{Observable features of the excitation spectrum}

The analysis of the above section reveals that the excitation spectrum is composed of two distinct sectors.
The first one is realized by the (finite) set of localized excitations with discrete eigenvalues.
Applying the terminology introduced in section \ref{SEopenSystem} to the present case, we can affirm that such localized excitations behave as
\emph{observable} quantities and the associated energies possess ideal features of reproducibility:
if the numerical basis employed is complete enough to
represent the states $\{\phi_p^a,\chi_p^a\}$, the energy $\Omega_a$
converges to its reference value.
A computational setup in which delocalized degrees of freedom are excluded by design can, in principle, describe excitations of this sector in an exact way.

The second sector contains a continuum spectrum of delocalized excitations. The associated states $\ket{\phi_p^a}$ behave as the unbound virtual orbitals analyzed in \cite{boffi2016} (see Sec.\ref{SEopenSystem})
and are deeply affected by the computational setup used to represent them. In particular, the presence of a finite computational domain implies a fictitious discretization of the excitation levels and
the associated spectrum exhibits an explicit dependence on the size of the domain.
Excitations belonging to this sector thus lose their intrinsic physical meaning and have to be understood simply as objects providing an effective representation of the density of states (DOS) of the Liouvillian superoperator, by means of the spectral operators $\sop B_a$ of Eq.~\eqref{spectralop}.

\begin{figure}[ht]
\includegraphics[scale=0.6]{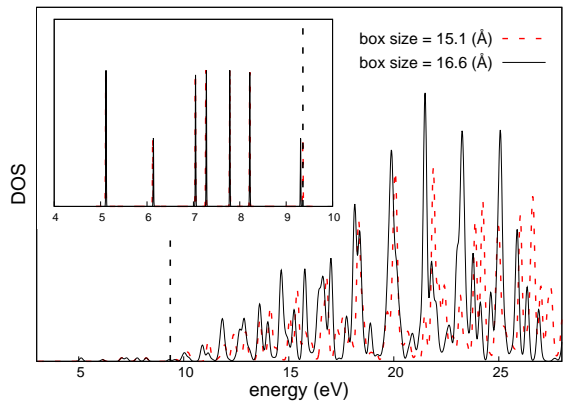}
\caption{(Color online) DOS of $CO$ excitation energies $\Omega_a$. The two curves correspond to two computational domains of different sizes. The inset contains the sole contribution of localized excitations. The first IP threshold is depicted as the vertical dashed line.}
\label{CO_exc}
\end{figure}
\begin{figure}[ht]
\centering
\subfloat[Excitation landscape]
{\label{c6h6_excLand}
\includegraphics[scale=0.56]{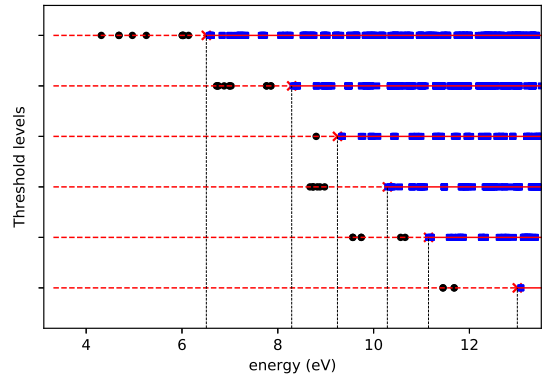}} \\
\centering
\subfloat[DOS of Excitation energies]
{\label{C6H6_exc}
\includegraphics[scale=0.58]{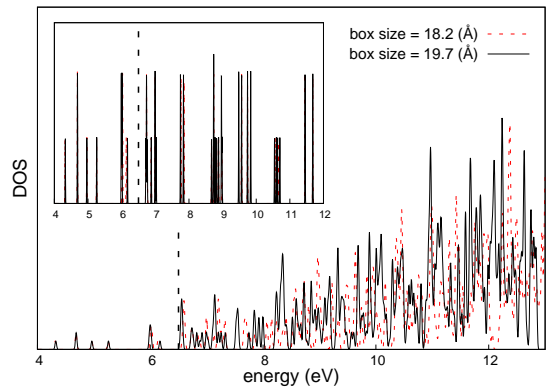}}
\caption{(Color online) Benzene molecule. (a): Excitation Landscape: the excitations have been separated into discrete (black circle) and continuum (blue square) sectors following the criteria of Appendix~\ref{ExcitationLandscape}. (b): DOS of excitation energies $\Omega_a$. The first IP threshold is depicted as the vertical dashed line.}
\end{figure}

In Fig.~\ref{CO_exc} and \ref{C6H6_exc} we plot the DOS of the Liouvillian for the $CO$ and benzene molecules,
for two different numerical setups of increasing simulation domain size.
The figures show clearly that all the excitations of energy $\Omega_a$ lying below the
molecule ionization potential (IP) have a localized behavior, as their energy does not vary with the size of the simulation domain.
On the other hand, above the first ionization threshold the excitations have an energy that strongly depends on the computational
setup, which is in line with the expected pseudo-continuum character of this sector.
In addition, for the benzene molecule it is also possible to
identify excitations of energy above IP that still possess a localized behavior, fulfilling the condition $\Omega_a < |\eps_a|$ [see the inset of
Fig.~\ref{C6H6_exc}]. In Fig.~\ref{c6h6_excLand} we collect the excitations of the benzene molecule following this criterion:
the agreement with the expected analytic structure presented in Fig.~\ref{ExcitationLandscape} is remarkable.

\subsection{Relative importance of discrete and continuum excitations}

So far, we have showed with both formal arguments and numerical calculations which show that the excitation spectrum splits into two sectors with very distinct features.
It is then reasonable to ask which is their relative importance and, in particular, if there is the possibility to express observable quantities, in a suitable $\omega$ range, by only employing the sector of genuinely localized excitations.
Stated otherwise, can the \emph{restriction} of $\sop \chi(\omega)$ to only localized excitations be useful to extract some physical quantities?

In this regard it is illustrative to apply the above arguments to the evaluation of the dynamical polarizability tensor $\alpha_{ij}(\omega)$.
Employing the formalism of Eq.~\eqref{deltaoexc} provides
\be\lb{imalpha}
\mathrm{Im}\left(\alpha_{ij}(\omega) \right) =
\sum_{a} \trace{\mathbf r_i \sop B_a \mathbf r_j} \delta(\omega - \Omega_a) \;.
\ee
The imaginary part of the dynamical polarizability tensor therefore can be seen as the partial density of states of the Liouvillian projected on
$\op{\mathbf r}_i\op{\mathbf r}_j$. For this reason, it is evident that below the \emph{first} excitation threshold (IP), the contributions to $\mathrm{Im}\left(\alpha_{ij}(\omega)\right)$ can be expressed only in terms of localized, thus observable, excited states. On the contrary, the non-observable, basis dependent continuum sector of the excitation starts to contribute to $\mathrm{Im}\left(\alpha_{ij}(\omega)\right)$ above IP; the localized sector is not anymore sufficient to express Eq.~\eqref{imalpha} in this $\omega$ range.

To further address this point, we exploit the formal relation between response density and susceptibility functional $\dm'(\omega) = \sop \chi(\omega)\op\Phi(\omega)$ and make usage of the representation ~\eqref{rhoPrimeFluctuationStateDef1} of the response density.
Pursuing this procedure provides the excitation-based representation of FS as follows:
\be\lb{fsinexc}
\ket{f_p^{\mathbf r_j}(\omega)}=
\sum_a \frac{\bra{w_p^a} \op{\mathbf r}_j \ket{\psi_p}}{\omega^2 - \Omega_a^2}
\ket{w_p^a}\;,
\ee
where we have employed the notation $\ket{w_p^a}=\ket{\phi_p^a}+\ket{\chi_p^a}$.
This is the well-known equivalence of the Sternheimer formalism with the spectral representation of the LR susceptibility, that is always valid
in any computational basis set provided the the full spectrum of excited states (including the pseudo-continuum ones) is considered.
The above equation shows that both the sectors of localized and delocalized states $\ket{w_p^a}$ contribute to the FS for \emph{any} value of $\omega$.
However, the discussion of Sec.~\ref{FluctuationState} has shown that the FS is a genuinely localized state below the IP threshold.
The equivalence implies that in this regime the contribution of the delocalized excitations in Eq.~\eqref{fsinexc} has to resum into a localized, asymptotically vanishing FS.
The localization below threshold is manifest for the imaginary part of the FS, that is explicitly written as the sum over $\ket{w_p^a}$ below IP [see Eq.~\eqref{imalpha}].
Instead, for the real part of FS, the \emph{entire} excitation spectrum contributes to the construction of the FS,
which has to be interpreted as a localized wave-packet in the basis of the pseudo-continuum excitations.

Hence, contrary to GS, we conclude that \emph{for LR calculations, even when the response density is localized, the contribution coming from the
continuum sector of the excitations can never be neglected}. In other terms, we cannot exclude that a fluctuation state, even if genuinely localized, will have a nonzero projection into the essential part of the Liouvillian spectrum, which in turn cannot be expressed by only employing localized states.
This fact shows that it is \emph{impossible} to restrict the spectral representation of $\chi(\omega)$ to only localized, discrete, \emph{code-independent} excitations, even for representing LR results in the $\omega=0$ limit. The basis-dependent sector of the excitation will be needed to
represent the linear susceptibility in the appropriate region of space.

\subsection{Effective representation of $\sop \chi(\omega)$ for localized response densities}

The above arguments have shown that the only quantity which can be expressed in terms of the localized sector of the excitation spectrum is the imaginary part of the susceptibility functional $\sop \chi(\omega)$ in the optical regime (i.e. $\omega <$  IP). The response density for a generic perturbation and $\omega$ value requires the inclusion in the Liouvillian DOS of the delocalized part of the excitation spectrum, even in view of expressing localized fluctuation states.

A given numerical treatment typically provides only an effective description of the sector of delocalized excitation which in turns gives rise to an effective
representation of the susceptibility functional. It appears therefore important to identify indicators that enable us to validate the capability
of $\sop \chi(\omega)$ to express the localized FS [and thus the complete response density $\dm'_\Phi(\omega)$] below the IP threshold.
For instance, the static polarizability tensor $\alpha_{ij}$ can be calculated as
\be\lb{sminus2}
\alpha_{ij} = \sint \dd a \frac{w^a_{ij}}{\Omega_a^2}\,, \quad w^a_{ij} \equiv \sum_{p\in a} |\bra{\psi_p} \op {\mathbf r}_i \ket{w_p^a}|^2 \;.
\ee
Such a value can be straightforwardly compared with results of Eq.~\eqref{staticalpha}, where the localized FS is obtained from GS techniques as
described in Sec.~\ref{FluctuationState}.
This is the well-known $S_{-2}$ sum-rule (see, e.g.,\cite{Wagner2012}), that expresses the value of the static polarizability
in terms of the molecule's oscillator strengths.
In the context of a numerical evaluation of the excitations, such sum rule can therefore be interpreted as the capability of the pseudo-continuum sector of the excitation to
express the localized FS -- and consequently the reference value of $\alpha_{ij}$, at $\omega=0$.

\begin{figure}
\centering
\subfloat[$CO$]
{\label{CO_alphavsExc}
\includegraphics[scale=0.6]{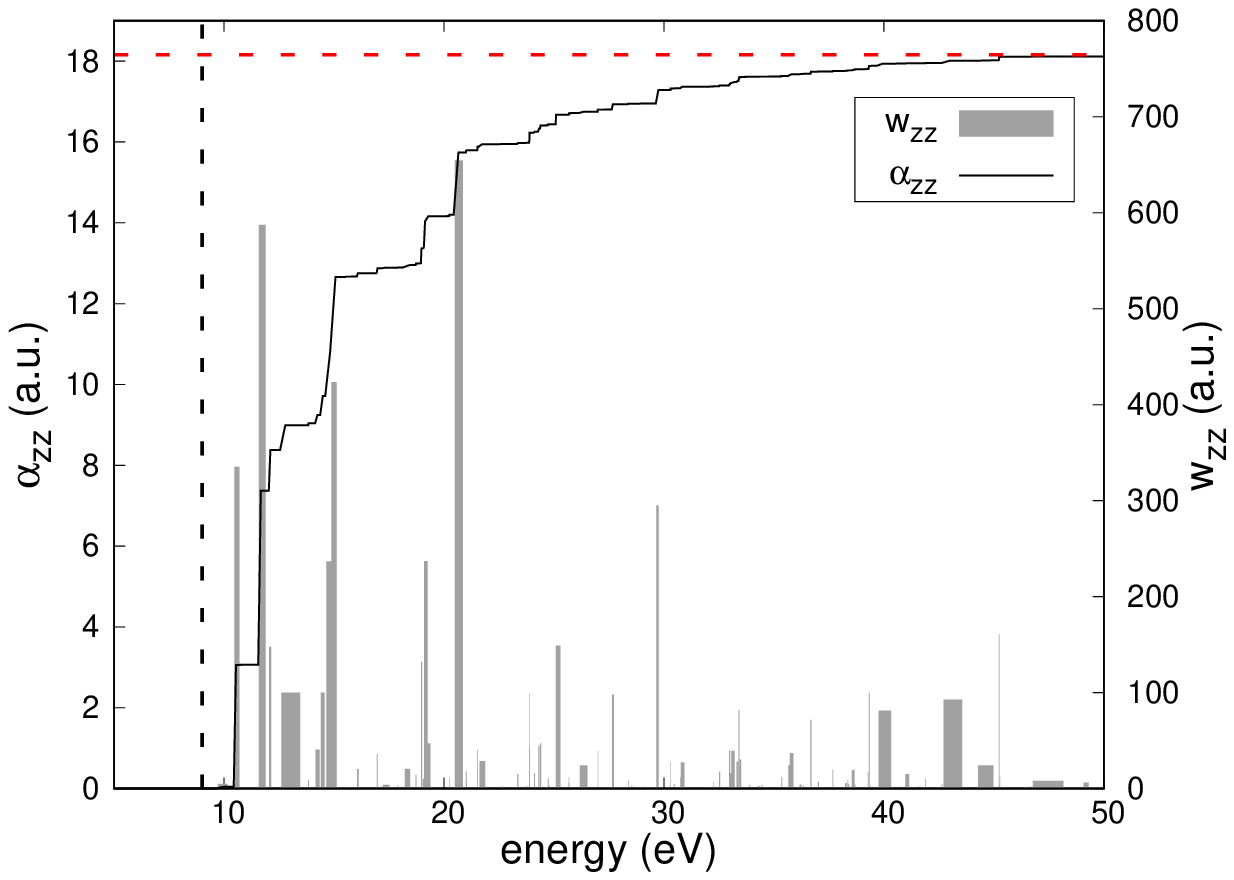}} \\
\centering
\subfloat[$C_{60}$]
{\label{C60_alphavsExc}
\includegraphics[scale=0.6]{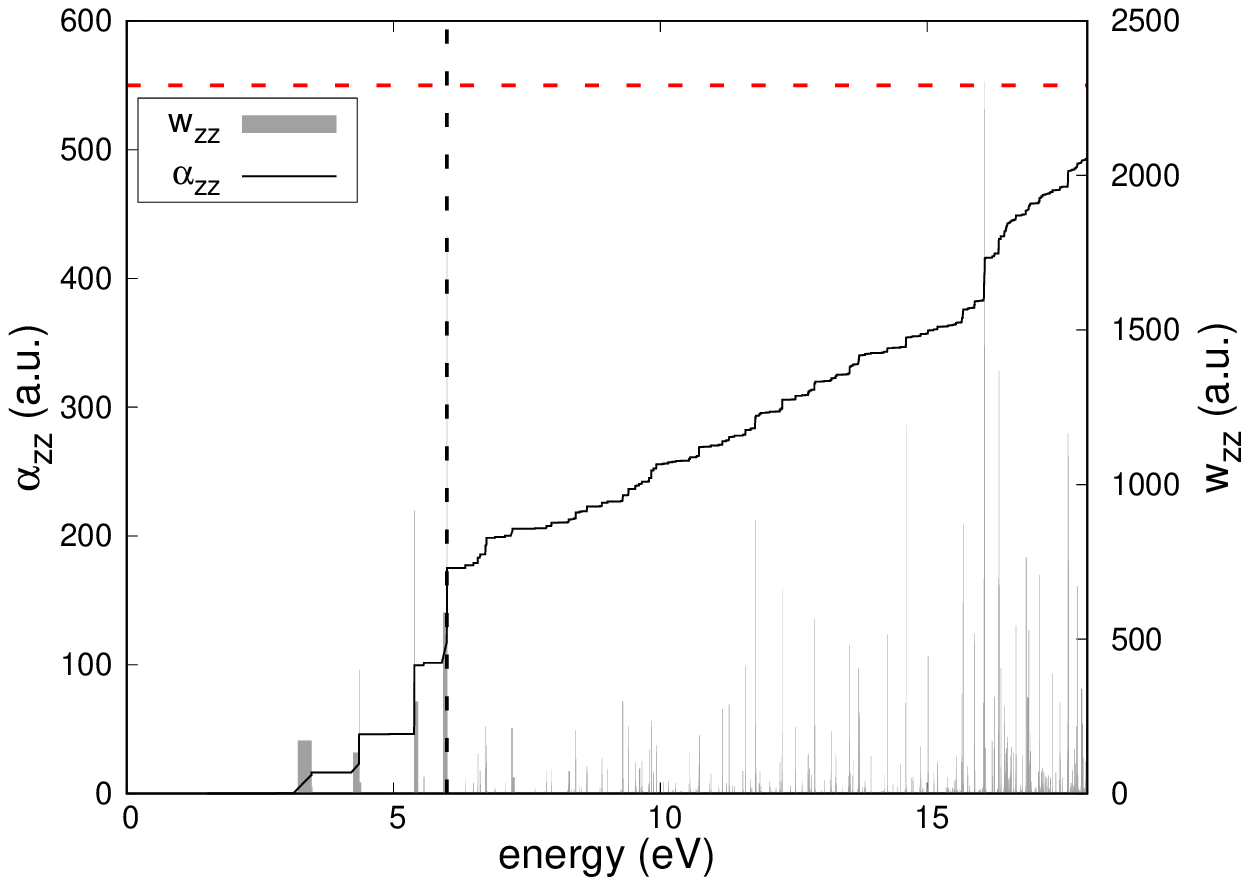}}
\caption{\label{co_AlphaExc}(Color online) Convergence of the static polarizability $\alpha_{zz}$ of the $CO$ (panel a) and $C_{60}$ (panel b) molecules,
w.r.t the excitations considered in the integral of Eq.~\eqref{sminus2}. The reference value obtained from Eq.~\eqref{staticalpha} is represented as a
horizontal red dashed line. The value of the IP energy is indicated by a vertical black dashed line. For each excitation energy, the values of the
oscillator strengths $w^a_{zz}$ of Eq.~\eqref{sminus2} are also plotted. }
\end{figure}

We have analyzed the fulfillment of Eq.~\eqref{sminus2} for molecules of various
sizes and symmetries.
In Fig.~\ref{co_AlphaExc} we show explicitly such a comparison for the $CO$ and $C_{60}$ molecules.
We notice that, in both the cases, a considerable number of pseudo-continuum excitations is required
to find the correct result; the sector of the discrete excitations appears to be largely insufficient to express the
reference fluctuation state.
In particular, in the case of $CO$, the complete excitation set is able to reproduce the reference value of $\alpha_{zz}$, but the contribution of localized sector is negligible w.r.t. the one coming from the continuum one. Instead, for the $C_{60}$, the localized sector provides a more significant contribution but a higher number of the psuedo-continuum excitations should be included in the sum of \eqref{sminus2} to satisfy the $S_{-2}$ sum-rule.

\begin{table}
\begin{tabular}{c|cc||cc}
  Mol. ($N_{at}$) & $\braket{\alpha}_\text{ref}$ & $ \braket{\text{Eq.~\eqref{sminus2}}}$ & \% LS & \% $\Omega_a <$ IP  \\
  \hline
  \hline
  N$_2 (2)$ &12.49 & 12.38 & $1.6\cdot 10^{-6}$ & $1.6\cdot 10^{-6}$ \\
  CO (2)& 15.07 & 14.79 & 8.6 & 8.6 \\
  H$_2$O (3) & 10.87 & 10.72 & 0.1 & 0.1 \\
  Benzene (12)& 71.11 & 70.32 & 3.0 & $7.2 \cdot 10^{-5}$\\
  Aflatoxin (35)& N/A   & 226.27 & 15.87 & 14.84 \\
  C$_{60}$ (60)& 561.59 & 509.60 & 38.47 & 34.41
\end{tabular}\caption{Comparison of the average static polarizabilities for molecules of various sizes (indicated by the number of atoms $N_{at}$) and symmetries, obtained from the reference calculations of Eq.~\eqref{staticalpha} and from the sum over the excitations extracted from the Casida coupling matrix. The last two columns show the percentage of the contribution to Eq.~\eqref{sminus2} which
come from the localized sector (LS) of the excitations, by focusing also on the LS before the first IP, where all the excited states are localized. We see that by increasing the molecules' sizes the LS contribution increases in percentage, however being always largely insufficient; the contribution coming from the pseudocontinuum excitations can never be neglected.}
\label{alphaTable}
\end{table}

This scenario is confirmed by the results for the other molecules presented in Table \ref{alphaTable}; varying the molecule size
and symmetry does not modify the qualitative behavior of the excitation landscape (see Supplementary Material at~\footnote{\texttt{URL inserted by the publisher}} for further details about the dependence of the excitation landscape on the size of the molecule).
In addition the localized sector of the excitations alone is largely insufficient to express $\alpha_{ij}$. Such considerations
are independent of the number of atoms of the molecule, as the range in $\omega$ needed to achieve satisfactory agreement is largely above the last valence ionization threshold, which is an intensive quantity.

Once again, this proves that, for any numerical treatment,  pseudo-continuum excitations are fundamental in view of a correct expression
of the response density below IP.
However, the fulfillment of $S_{-2}$ is not a universal quality indicator of $\chi(\omega)$ as the latter may be still badly represented in the high frequency regime:
it is enough to recall
the strong dependency on the box size of the absorption spectra of Fig.~\ref{CO_exc}.
For all these computational setups, the $S_{-2}$ sum rule is, on the contrary, well satisfied.

\section{Conclusions and Perspectives}

We have performed a critical assessment of the fundamental equations that govern the linear response theory of molecular systems.
We have shown that the response density for values of $\omega$ above the first IP \emph{has} to be expressed in terms of genuinely delocalized states.
This result has relevant consequences in the choice of the  appropriate computational basis set to be employed for the LR treatment.

For frequencies below the IP threshold, under reasonable assumptions on the perturbing operators, the response density is genuinely localized.
In this regime, it is possible to achieve convergence of the numerical result by increasing the basis completeness in the same spirit as GS calculations.
More complete basis sets will provide more precise results, regardless of their asymptotic behavior. A comparison among various computer codes in view of reproducibility should therefore be performed initially in this regime.

From a computational perspective, a set of localized basis functions, which \emph{by design} excludes long-range oscillatory behavior, may be complete
enough to express $\sop \chi(\omega)$ \emph{below} IP, but it will never be able to capture the entire features of the response density operator above
the ionization threshold: the system's excitations expressed in this basis are reliable \emph{only} in the optical regime.
High-energy excitations belong to a continuum sector (which should not be confused with the continuum eigenstates of the unperturbed GS Hamiltonian)
and the excitation density of states for $\omega >$ IP provided in this way has to be considered as an effective representation of $\sop \chi(\omega)$ in order to fulfill Eq.~\eqref{fsinexc}. This fact has been pointed out several times (see, e.g., \cite{giustino2012,giustino2014}) in the literature.

On the contrary, computational treatments that are able to express delocalized states may be adequate to describe more efficiently the excitations DOS,
especially due to the capability of the basis to capture oscillatory behavior. Like in localized basis sets, excitations below threshold can be
precisely calculated with the same paradigm of traditional GS calculations. However, in this computational setup, to correctly express the linear susceptibility below IP -- more specifically the real part of the fluctuation state -- care should be taken in considering enough delocalized excitations
to guarantee fulfillment of sum-rules like Eq.~\eqref{sminus2}.

For generic LR quantities, which are functionals of the linear susceptibility, the quality of $\sop \chi(\omega)$ will then depend
on the frequency $\omega$ of interest: a given spectral representation may provide high-quality results for static $\omega=0$ regimes, while being
unable to provide convergence for high-energy absorption spectra.
Absorption spectra below IP are relatively easy to converge, as they are observables which depend only on localized quantities.
However, the fulfillment of sum rules like the $S_{-2}$ is not \textit{per se}
a guarantee of the quality of high energy absorption spectra.
Moreover, if the basis set employed for the evaluation of the target $\alpha_{ij}$ in  Eq.~\eqref{sminus2} is identical to the one employed for the LR treatment, the $S_{-2}$
is trivially satisfied in the basis. It is therefore important to have a set of reference values
in the complete basis set limit.

We believe that these considerations, based on simple manipulations on the fundamental LR equations, will help to increase the reliability of LR
calculations in the community, and in building test-sets that would help in the \emph{calibration} of present and future computer codes, thereby
increasing the predictivity of present-day theoretical approaches. The example of the static polarizability of molecules
presents ideal features as an initial playground, as it has already been pointed out recently~\cite{HeadGordon}. Work is ongoing in this direction.

\begin{acknowledgments}
We thank Thierry Deutsch for useful comments on the manuscript and Laura Ratcliff for proofreading.
\end{acknowledgments}

\appendix
\section{Transverse operators and right action of the Liouvillian superoperator}\lb{LiouvillianAction}
Given an unperturbed Hamiltonian and density operator, which satisfy $\commutator{\hnot}{\dmnot}=0$, we may decompose a generic operator in two contributions, $\op O=\op O_\parallel + \op O_\perp$, defined as:
\begin{align}
\op O_\parallel &\equiv \dmnot \op O \dmnot + \op Q_0 \op O \op Q_0 \nn \;,\\
\op O_\perp &\equiv \dmnot \op O \op Q_0 +  \op Q_0 \op O \dmnot =
\commutator{\dmnot}{\commutator{\dmnot}{\op O}} \;,
\end{align}
with $\op Q_0 =\identity - \dmnot$.
The operator $\op O_\parallel$ is constructed to satisfy
$\commutator{\dmnot}{\op O_\parallel} = \commutator{\hnot}{\op O_\parallel} =0$.
Also, it is easy to verify that given two generic $\op O$, $\op O'$, we have
$\trace{\op O'_\parallel \op O_\perp} =0$.

The projection $\op O_\perp$ is therefore the relevant part for the commutator
\begin{equation}
\commutator{\dmnot}{\op O} = \commutator{\dmnot}{\op O_\perp} =
\dmnot \op O \op Q_0  - \op Q_0 \op O \dmnot  \;,
\end{equation}
consequently the Liouvillian superoperator, which contains by definition a commutator with $\hnot$ and $\dmnot$ in its unperturbed and coupling part respectively,
is constructed such as $\Liouv \op O = \Liouv O_\perp$. In addition, the image operator satisfies the transverse condition
$\left( \Liouv \op O \right)_\parallel =0$.

Here $\op V'$, which expresses the perturbation induced by the density dependence on ground state Hamiltonian, can be conveniently expressed by defining the
\emph{scalar coupling kernel}
\be\lb{CouplingKernelDef1}
U\left[\op O; \op O'\right] \equiv  \int \dd \r \dd \r' \trace{\op O \frac{\delta \op V[\dmnot] }{\delta \rho(\r,\r')}
} \trace{\op O' \ket{\r} \bra{\r'}}\;,
\ee
on the basis of which
\be\lb{VpDef1}
\op V'[\op O]=
\int \dd \r \dd \r' U\left[\ketbra{\r}{\r'};\op O\right] \ketbra{\r'}{\r}\;.
\ee
We notice here in passing that in general $\op V'[\op O_\perp] \neq \op V'[\op O]$.
We assume nonetheless that the scalar coupling kernel is symmetric, i.e. $U\left[\op O; \op O'\right] = U\left[\op O'; \op O\right]$, which is a condition generally satisfied in DFT Hamiltonians.

The action from the right of $\Liouv$ can be assessed through the equivalence $\trace{\op O'(\Liouv\op O)} = \trace{(\op O'\Liouv)\op O_\perp}$, which implies that also
$\left(\op O' \Liouv\right)_\parallel =0$. We obtain:
\be
\op O\Liouv = -\commutator{\hnot}{\op O} + \int \dd\r\dd\r'
U\left[\commutator{\dmnot}{\op O};\ketbra{\r'}{\r}\right]\left(\ketbra{\r}{\r'}\right)_\perp \;. \nn
\ee
With this definition it is explicitly apparent that $\op O\Liouv=\op O_\perp \Liouv$,
similarly to the left action. The action of the Liouvillian reads as
\begin{align}\lb{LiouvillianRightActionDef1}
\op O\Liouv &= -\commutator{\hnot}{\op O} + \left(V'\left[\commutator{{\dmnot}}{\op O}\right]\right)_\perp = -\commutator{\hnot}{\op O} + \nn \\
&+ \dmnot\op V'\left[\commutator{{\dmnot}}{\op O}\right]\op Q_0 + \op Q_0\op V'\left[\commutator{\dmnot}{\op O}\right]\dmnot \;.
\end{align}
In the last equivalence we have made usage of the transverse property [see Eq. \eqref{RhopTransverseDef1}] of the coupling superoperator. The above formulas
together with Eq.~\eqref{LiouZeroDef1} are the starting point for the solution of the eigenvalues equations \eqref{ExcitationOperatorsDef1}.

\section{Casida equations}
\label{casida}

We show that Casida equations are equivalent to the equations of motion for the excited states \eqref{ExcitationOperatorsDef3}. To this aim we introduce an explicit basis $\{\ket{s}\}$
in the subspace of empty states so that both $\ket{\phi^a_p}$ and $\bra{\chi^a_p}$ can be represented as
\be
\ket{\phi^a_p} = \sum_s X^a_{p s}\ket{s} \;, \qq
\bra{\chi^a_p} = \sum_s\bra{s}Y^a_{p s} \;. \lb{phiandchi}
\ee
Plugging this expansion into Eqs. \eqref{ExcitationOperatorsDef3} and projecting on a arbitrary element of the basis provides a linear systems of equations for the coefficients
$X^a_{ps}$ and $Y^a_{ps}$, namely
\begin{align}
 \sum_{s'}(H_{0ss'}-\eps_p\delta_{ss'})X^a_{ps'} + \bra{s}\op V'[\op E_a]\ket{\psi_p} &= \Omega_a X^a_{ps} \;, \nn \\
 -\sum_{s'} (H_{0 s' s}-\eps_p\delta_{s s'})Y^a_{ps'} - \bra{\psi_p}\op V'[\op E_a]\ket{s}& = \Omega_a Y^a_{ps} \;. \nn
\end{align}
Using this basis, excitation operators are expressed as a linear combination of transition operators $\excite{p}{s} \equiv \ketbra{s}{\psi_p}$ and $\decay{s}{p} \equiv \ketbra{\psi_p}{s}$,
as follows
\be\lb{ExcitationOpBasisTransition1}
\op E_a = \sum_{ps}\excite{p}{s}X^a_{ps}+\decay{s}{p}Y^a_{ps} \;.
\ee
The contribution of the coupling operator in the equations of motion of excited states can be expressed by means of the coupling kernel \eqref{CouplingKernelDef1} as
\begin{align}
& \bra{s}\op V'[\op E_a]\ket{\psi_p} = \sum_{qs'}\left(U[\decay{s}{p};\excite{q}{s'}]X^a_{qs'} + U[\decay{s}{p};\decay{s'}{q}]Y^a_{qs'}\right) \;,\nn \\
& \bra{\psi_p}\op V'[\op E_a]\ket{s} = \sum_{qs'}\left(U[\excite{p}{s};\excite{q}{s'}]X^a_{qs'} + U[\excite{p}{s};\decay{s'}{q}]Y^a_{qs'}\right) \;, \nn
\end{align}
Given that we employ real functions both for $p$ and $s$ we also have $\excite{p}{s}(\r,\r') = \decay{s}{p}(\r',\r)$ and the density operator is symmetric in $\r \leftrightarrow \r'$.
These conditions applied together imply that $U[\excite{p}{s};\excite{q}{s'}] = U[\decay{s}{p};\decay{s'}{q}]$, as well as $U[\decay{s}{p};\excite{q}{s'}] = U[\excite{p}{s};\decay{s'}{q}]$,
and so Eqs. \eqref{ExcitationOperatorsDef3} can be recasted in matrix form
\be\lb{ExcitationMatrixEq1}
\sum_{qs'}\mat{F_{ps}^{qs'} &  D_{ps}^{qs'}  \\
- D_{ps}^{qs'} & - F_{ps}^{qs'} }
\mat{X^a_{qs'} \\ Y^a_{qs'}} = \Omega_a \mat{X^a_{ps} \\ Y^a_{ps}}
\ee
where
\begin{align}
& F_{ps}^{qs'} = (H_{0 ss'}-\eps_p\delta_{ss'})\delta_{pq} + U[\decay{s}{p};\excite{q}{s'}] \;,\nn \\
& D_{ps}^{qs'} = U[\decay{s}{p};\decay{s'}{q}] \;.
\end{align}
This is the well-known Casida equation in the basis of transitions for the excitation of energy $\Omega_a$. The computational reliability of the
eigenvalue $\Omega_a$ depends on the capability of the basis set $\ket{s}$ to fulfill of Eq.~\eqref{phiandchi}. Note that this may be valid only for a subset of the solutions of the eigenproblem \eqref{ExcitationMatrixEq1}.

\section{Computational details} \lb{compdetails}
The illustrative calculations presented in this paper are all KS-DFT calculations at the LDA level of theory. We employed the BIGDFT code \cite{BigDFT},
that makes usage of Daubechies Wavelets as computational basis set. This orthonormal basis presents optimal features in view of reproducibility of the results, and it has proved to represent precisely and explicitly a system with free as well as periodic boundary conditions (BC), without any basis set superposition error nor supercell aliasing, for free BC.

For ground-state results, the important parameters in the context of convergence of the calculations are the spacing of the wavelet grid and the size of the simulation domain: for bound-state like functions, convergence is achieved by reducing the grid spacing and increasing the simulation box size.
Norm-conserving pseudopotentials (PSP) of the HGH \cite{NLCCPSP} type are employed to remove the core electrons. Such PSP have proved to be able to provide all-electron accuracy for molecular systems.

Our LR calculation is performed with the Casida formalism \cite{bhaarathi}, where it is also important to include a sufficient number of unoccupied states in the transition operators. As pointed out in the text such states behave as plane waves: the features of wavelets enable us to represent localized and delocalized states on equal footing.
For each size of the simulation domain, the number of unoccupied states used for building the Casida' eigenproblem has been increased up to convergence of the presented results.

All the details of the calculations, the \textsl{Jupyter} notebooks of the data analysis, including the ones presented in the Supplementary Material, may be found at the URL~\footnote{\protect\url{https://github.com/luigigenovese/LR-nb/blob/master/POLARIZABILITY/nb_paper/LR_Analysis.ipynb}}

\subsection{$CO$}
The molecule is oriented along the $z$ axis and the $z$-dimension of the box is equal to 11.5, 15.1 and 16.6 \AA.
Bond length is 1.235\AA, stretched w.r.t. the equilibrium LDA position such as to increase the value of the polarizability.
280 unoccupied states have been computed in each setup, which span an energy range of the empty KS orbitals of 30.7, 19.3 and 15.4 eV respectively.
These energy values should not be confused with the excitations energy ranges reported in Figs.~\ref{co_spectrum} and \ref{CO_exc}.
Once again, such choices of unoccupied states ensures convergence of the excitation spectrum in the presented range.

The DOS reported in figure \ref{CO_exc} are obtained by adding a smearing parameter of $8\times 10^{-2}$ eV whereas in the inset the shift is $5\times 10^{-3}$ eV.

\subsection{Benzene}
The molecule is oriented in the $xy$ plain and the $x$-dimension of the box is equal to 15.0, 18.2 and 19.7 \AA.
220 states have been computed in each setup and the corresponding
energy range for the unoccupied states is 16.9, 11.6 and 9.7 eV.

The DOS reported in Fig. \ref{C6H6_exc} are obtained by adding a complex shift of $1.3\times 10^{-2}$ eV whereas in the inset the shift is $5\times 10^{-3}$ eV.

\subsection{Excitation Landscapes}
The attribution of each excitation of the appropriate sector passes through the evaluation of the norm of the states $\phi_p^a$ and $\chi_p^a$, in the form of Eq.~\eqref{phiandchi}. Only states with norm greater than a given tolerance value(taken as the 1\% of the total norm) are considered and the associated
excitation is attributed to the first sector if the condition $\Omega_a < |\epsilon_p|$ with $\epsilon_a = \mathrm{max}\left(\epsilon_p\right)_{p\in a}$.
This computational procedure provides results in optimum agreement with the formal arguments reported above. Indeed the excitation landscape of
[for instance, Fig.~\eqref{c6h6_excLand} for the case of benzene] evidences the presence of discrete excitations (the black bullets) embedded in a pseudo-continuum
represented by the blue filled square. The comparison between the excitation DOS computed in different computational domain shows that this way of labeling the excitation is in line with the locality of the corresponding states since the corresponding energies easily converge with respect to the increase of the simulation domain.

\bibliography{Analytic_biblio}

\end{document}